# Breaking the challenge of signal integrity using time-domain spoof surface plasmon polaritons

Hao Chi Zhang[1,2,*], Tie Jun Cui[1,3,*,†], Qian Zhang[1,2], Yifeng Fan[1,2], and Xiaojian Fu[1,2]

[1] *State Key Laboratory of Millimeter Waves, Southeast University, Nanjing 210096, China*
[2] *Synergetic Innovation Center of Wireless Communication Technology, Southeast University, Nanjing 210096, China*
[3] *Cooperative Innovation Centre of Terahertz Science, No.4, Section 2, North Jianshe Road, Chengdu 610054, China*

*These authors contributed equally to this work.
†To whom correspondence should be addressed. E-mail: tjcui@seu.edu.cn

**In modern integrated circuits and wireless communication systems/devices, three key features need to be solved simultaneously to reach higher performance and more compact size: signal integrity, interference suppression, and miniaturization. However, the above-mentioned requests are almost contradictory using the traditional techniques. To overcome this challenge, here we propose time-domain spoof surface plasmon polaritons (SPPs) as the carrier of signals. By designing a special plasmonic waveguide constructed by printing two narrow corrugated metallic strips on the top and bottom surfaces of a dielectric substrate with mirror symmetry, we show that spoof SPPs are supported from very low frequency to the cutoff frequency with strong subwavelength effects, which can be converted to the time-domain SPPs. When two such plasmonic waveguides are tightly packed with deep-subwavelength separation, which commonly happens in the integrated circuits and wireless communications due to limited space, we demonstrate theoretically and experimentally that SPP signals on such two plasmonic waveguides have better propagation performance and much less mutual coupling than the conventional signals on two traditional microstrip lines with the same size and separation. Hence the proposed method can achieve significant interference suppression in very compact space, providing a potential solution to break the challenge of signal integrity.**

# INTRODUCTION

The signal interference suppression and signal integrity are one of the most challenging topics in physics and electrical engineering. Especially, the rapid developments of super-large-scale integration of high-speed circuits and wireless communication systems/devices have brought forward higher requirements to the signal interference suppression and signal integrity in the past decades *(1)*. Some special circuit strategies have been proposed to improve the signal quality, such as the equalization technique *(2)* and differential microstrip lines *(3)*. However, in above occasions, additional areas of circuits and power consumptions are required, which make them rather difficult to be utilized in the high-frequency and/or high-speed circuits and systems. In fact, the signal integrity, signal interference suppression, and miniaturization are three key features to be solved simultaneously to achieve higher performance and smaller size of complicated circuits and systems. But these factors are almost contradictory using the current techniques of transmission lines. To solve the challenges, we are forced to find new methods and techniques, in which the surface plasmon polariton (SPP) approach is one of the possibilities due to its strong subwavelength effects.

At optical frequencies, because of the negative permittivity behavior of metal and positive permittivity of dielectric, SPPs are formed by the interaction between free electrons and the electromagnetic field *(4, 5)*, and propagate in parallel to the metal-dielectric interface with the exponential decay in the direction perpendicular to the interface. These special surface waves have intrigued great interests and been studied intensively during the past decades, owing to the extraordinary talent of field confinement *(6)* and potential applications in super-resolution imaging *(7, 8)*, miniaturized sensors *(9)*, and photo voltaic *(10)*. On account of the unique properties of SPPs, the optical circuits based on SPPs are generally recognized as one of the solidest venue for the further developments, although some sophisticated theories still need to be established and validated *(11)*.

Even though SPPs have such excellent superiorities, they cannot be naturally achieved at low frequencies (i.e. microwave and terahertz) because metals show the characteristics of perfectly electric conductor (PEC) rather than plasmon with the negative permittivity. To overcome this difficulty, the idea of constructing SPPs using metamaterials, so-called as spoof

SPPs or designer SPPs *(12)*, has been proposed *(13)*. Since spoof SPPs inherit the properties of SPPs and metamaterials, they can realize most of the exotic features of optical SPPs, and the physical characteristics of spoof SPPs can be designed at will by tuning the geometrical parameters *(14-21)*. More recently, an ultrathin corrugated metallic strip structure *(22)* has been reported to realize the conformal surface plasmons (CSPs), which is one of the most potential candidates for the conformal circuits on account of their flexibility and near-zero thickness. Based on the ultrathin corrugated metallic strip, an active spoof SPP device – the SPP amplifier – has been presented in the wideband microwave frequencies *(23)*, which can amplify the SPP waves by 20 dB, providing the foundation of potential spoof SPP integrated circuits and systems.

Here, we propose to solve the signal-integrity challenge using time-domain spoof SPPs as the carrier of signals in the microwave frequency. We design a special plasmonic waveguide that is composed of two ultrathin corrugated metallic strips printed on the top and bottom surfaces of a dielectric substrate with mirror symmetry *(23)*. Such a plasmonic waveguide is capable of supporting spoof SPPs from very low frequency to cutoff frequency, and hence time-domain spoof SPPs. When SPP signals are propagating along two very-closely-packed plasmonic waveguides, as shown in Fig. 1A, we illustrate excellent transmission performance of SPPs and powerful ability to suppress the signal interference, which are verified by both numerical simulations and experiments. Possessing the features of strong field confinements, significant interference suppression in very compact space, and natural filtration, the proposed method may play an important role in realizing complicated SPP integrated circuits and systems.

## RESULTS AND DISCUSSION

The ultrathin conformal SPP structure is a big advance to the microwave and terahertz SPP circuits due to its flexibility and broadband *(22)*. But this structure cannot naturally coalesce with the conventional microwave circuits because of the single-line geometry. It was shown that the ultrathin corrugated metallic strip is able to confine the electromagnetic fields in two orthogonal directions and propagate the SPP waves with very low loss. However, the single line structure has two inevitable flaws in applications: 1) it is inconvenient to be integrated with an active chip in the microwave frequency *(23)*; 2) it cannot support the low-frequency SPP waves, which leads to signal distortion when transmitting time-domain signals.

It is well known that most traditional transmission lines are based on double-conductor or multi-conductor structures to support low-frequency electromagnetic wave propagation, which is essential for the time-domain signal transmission. Hence, we adopt here a special double-conductor plasmonic waveguide to support time-domain spoof SPPs in the microwave frequency. The plasmonic waveguide is structured by two corrugated metallic strips that are printed anti-symmetrically on the top and bottom surfaces of a dielectric layer *(23)*, as shown in Fig. 1C. The plasmonic waveguide structure is constructed by arranging mirror duplicated unit cells periodically along the *x* axis, in which the groove depth and groove width are denoted as *d* and *a*, and the strip width and thickness are *h* and *t*, respectively, with a period of *p*. Using flexible dielectric substrate, this kind of plasmonic waveguide can also be designed to support conformal SPP mode, if required.

Fig. 2A depicts the dispersion curves of the plasmonic waveguide with different groove depths *d*, in which the other geometrical parameters are chosen as *a*=0.96 mm, *h*=2.4 mm, *t*=0.018 mm, and *p*=2.4 mm, so that microwave SPPs are generated. The metal is selected as cooper, which can be regarded as PEC in the microwave frequency, and the dielectric substrate is Rogers RT5880 with dielectric constant $\varepsilon_r = 2.2$, loss tangent $\tan\delta = 0.0009$, and thickness *ts*=0.508 mm. According to the previous theory and experience *(24, 25)*, the ignored metal loss leads to a tiny blue shift of dispersion spectrum but shows negligible impact on the shape of dispersion curve. This will simplify the Eigen-mode simulation carried out by the commercial software, CST Microwave Studio, and avoid the complex calculations. In Fig. 2A, we clearly observe that the dispersion curves with different groove depths behave like the natural SPPs at optical frequencies. All curves are deviating gradually from that of the microstrip line, the conventional microwave transmission line, and further from the light line, and then asymptotically approach to different cutoff frequencies. More importantly, the performance of slow wave becomes more and more striking as the groove depth *d* increases from 0.4 to 2.0 mm. Therefore, the dispersion curves are sufficiently sensitive to the change of groove depth *d*, which allows easy tune of wave momentum, making it possible to realize a smooth conversion between the SPP waves and conventional guided waves.

One of the important features of the double-strip plasmonic waveguide is the guidance of low-frequency SPP waves, as demonstrated by the CST simulation results of transmission spectra in Fig. 2B, in which the plasmonic waveguide has been matched to the microstrip line with 50-ohm impedance by using the gradient corrugation grooves *(26-28)*. From Fig. 2B, we clearly observe that the transmission coefficients of the SPP waveguide are nearly unity (or 0dB) from 0 to 12 GHz, which has equivalent transmission performance to the microstrip line that has been widely used in modern integrated circuits and wireless communication systems. This important feature implies that time-domain SPP signals which cover a large range of frequency spectra can be guided by the plasmonic waveguide. Fig. 2C shows the transmitted time-domain SPP signal under the input of a Gaussian pulse with broad band from 0 to 12 GHz. We find that the transmitted SPP signal keeps very good waveform with tiny distortion, which is caused by the slight reflection and frequency dispersion.

Even though the frequency dispersion of the SPP waveguide produces a tiny distortion in broadband signal, the natural filtration provides an additional advantage to integrate circuits. As shown in Fig. 2A, each dispersion curve has the corresponding cut-off frequency. Once the operating frequency is higher than the cut-off frequency, the SPP waves cannot propagate through the plasmonic waveguide. This feature leads to a rapid decline in the transmission spectrum, as illustrated in Fig. 2B, which looks like a filter *(29, 30)*. This natural filtration provides a possibility to achieve miniaturizations of SPP circuits and systems by omitting the filter devices.

Owing to the interaction between such two corrugated metallic strips, the double-strip plasmonic waveguide demonstrates more significant subwavelength effect and tighter field confinement than the single corrugated strip under the same geometrical configuration, which helps reduce mutual coupling and miniaturize the SPP devices. To further show the spoof SPP propagations on the plasmonic waveguide, we give the electric-field distributions of the SPP modes with different phase shifts, as depicted in Fig. 3A. At low frequencies (smaller phase shifts), it is noted that the electromagnetic waves are mainly restricted between the two corrugated metallic strips, which is similar to the gap SPP mode in the coupling nanowire array *(31-34)*. As the frequency increases (i.e. the phase shift increases), we notice that the electric fields along the propagation direction are gradually enhanced, and the propagation mode gradually gets close to the SPP mode along single corrugated metallic strip *(22)*, which confines the electric fields around the surface structure. The mode change can be reasonably explained by the enhanced-confinement ability of the double-strip plasmonic waveguide with the increasing frequency.

In integrated circuits and wireless communication devices (e.g. cell phone) or systems, it is usually requested that many transmission lines are packed in a compact space, and the strong mutual coupling results in the signal-integrity problem. Here we propose to solve the problem by using the novel plasmonic waveguide and time-domain SPP signals. Without the loss of generality, we consider two closely packed plasmonic waveguides (see Fig. 1A) with a separation $s$=0.8 mm, which is about 0.027 wavelengths at 10 GHz. For easy comparison with the conventional technique, we also consider two closely packed microstrip lines with the same geometry and separation (see Fig. 1B). In fact, the physical insight for the plasmonic waveguide to reduce mutual coupling is the strong field confinement, which is also the major difference from the traditional microstrip line. This is confirmed by the numerical simulation results of electric fields under different states.

In Figs. 3B and C, we present the electric-field distributions on the plasmonic waveguides and microstrip lines on the *x-y* plane, which is 1.0 mm above the structure surface at 11 GHz, in which the fields are excited from the left of the lower SPP waveguide or microstrip line. We clearly observe that the electric-field energy is tightly localized in a very small region near the plasmonic structure, leading to no interference (or mutual coupling) to the adjacent upper waveguide. In the microstrip lines, however, a remarkable interference is observed, resulting in significant coupling fields propagating on the upper microstrip line. We further notice that the propagating wavelength on the plasmonic waveguide is much smaller than that on the microstrip line, which is convenient to miniaturize the SPP devices and circuits. To show the detailed distributions around the gap area of two SPP waveguides or microstrip lines, we provide the magnitude distributions of electric fields on the *y-z* plane, which is the central section of whole structure, as depicted in Figs. 3D and E. We clearly see that the overlap of the electric fields on SPP waveguides is much more inconspicuous than those of microstrip lines, which reduces the mutual coupling (or crosstalk) significantly.

To study the crosstalk reduction quantitatively, we apply the coupling theory to evaluate the coupling coefficient. According to the coupled-mode theory *(35)*, the transmitted power ratio (*T*) and coupling power ratio (*C*) of the adjacent waveguide system can be expressed analytically as

$$\begin{cases} T = \cos^2(\kappa L) e^{-2\alpha L} \\ C = \sin^2(\kappa L) e^{-2\alpha L} \end{cases} \quad (1)$$

in which $L$ is the coupling length of the waveguide, $\alpha$ is the imaginary part of the propagation constant, and $\kappa$ is the frequency-dependent coupling coefficient. For SPP modes, the coupling coefficient $\kappa$ is written as *(36)*:

$$\kappa = \frac{\omega \varepsilon_0}{4} \iint_{R^2} (n^2 - n_0^2) \left[ \mathbf{E}_{1t}^* \cdot \mathbf{E}_{2t} + \frac{n_0^2}{n^2} \mathbf{E}_{1n}^* \cdot \mathbf{E}_{2n} \right] dS \quad (2)$$

where $n$ and $n_0$ are the refractive indexes of surrounding medium and plasmonic waveguide, and $\mathbf{E}_{jt}$ and $\mathbf{E}_{jn}$ ($j$=1, 2) are normalized transverse and longitudinal electric fields of plasmonic waveguide $j$, respectively. From Eq. (2), we observe that the coupling of adjacent plasmonic waveguides is caused by the overlap of electric fields in such two plasmonic waveguides. Hence it is possible to suppress the crosstalk of two adjacent SPP waveguides using the excellent feature of field confinements of SPPs.

Since the SPP field confinements can also lead to field enhancements, which may cause an adverse result, the design of an appropriate separation interval is very important. Based on this consideration, we made numerical simulations of the transmission and coupling ratios of the plasmonic waveguides and microstrip lines with different intervals, as illustrated in Figs. 4A-D. From Fig. 4A, the curves with intervals *s*=0.4 mm have distinct dips in the transmission spectra of SPP waveguides, which suggests that the coupling of this case is potentiated by the field enhancement, and the behavior looks like a coupler rather than the crosstalk suppression. The curve with interval *s*=0.6 mm seems not to have a transmission dip, but in fact the dip caused by crosstalk meets the cutoff frequency, leading to a clearly different cutoff frequency. As the interval *s* increases from 0.8 to 2 mm, the transmission spectra do not exhibit noteworthy changes, which imply that the crosstalk is kept on a very low level in the passband (Fig. 4C). For traditional microstrip lines, however, the transmission performance becomes worse and the crosstalk becomes more significant monotonously as the interval *s* decreases, as shown in Figs. 4B and D.

In Eq. (2), we note that the coupling coefficient $\kappa$ is a function of the propagation constant. In traditional microstrip lines, the coupling coefficient gradually increases as the propagation constant increases (Fig. 4D) due to the difference of the squares of refractive index factor. But in allusion to the SPP waveguides, the increase of propagation constant leads to tighter field confinement, which reduces the factor of field-overlapping contribution. Hence, different from that in microstrip lines, the crosstalk in SPP waveguides appears undulation states in the frequency spectrum (Fig. 4C).

The small coupling coefficient and low loss of SPPs lead to another beneficial feature, i.e., the transmission power ratio is nearly independent of the coupling length *L*. This feature can be proved by the minor approximations of Eq. (1), which are written as

$$\begin{cases} T = \cos^2(\kappa L) e^{-2\alpha L} \approx 1 \\ C = \sin^2(\kappa L) e^{-2\alpha L} \approx \kappa^2 L^2 \end{cases} \quad (3)$$

Eq. (3) implies that the shape of the frequency spectrum is independent of the coupling length, which is confirmed by the numerical simulations shown in Figs. 4E and F. This feature is very convenient to the engineering design, which makes engineers do not consider the influence of the coupling length. The length-insensitive property of SPP waveguides is important evidence of the crosstalk suppression, which is completely different from the traditional microstrip lines.

For providing visualized evidence, the time-domain SPP transmissions along the two SPP waveguides are simulated, which are excited by two Gaussian-pulse inputs by a time interval of 0.5 ns, as shown in Fig. 5A. Here, with connections to potential applications, the input signal is chosen to have a bandwidth from 8 to 12 GHz, which is well known as the X band in the microwave frequency. From Fig. 5A, we note that the existence of adjacent plasmonic waveguide show negligible impacts on the transmitted SPP signals, which have nearly no distortions after propagating through the SPP waveguide, illustrating excellent transmission performance without interference. As a comparison, the traditional transmitted signals through the microstrip lines are demonstrated in Fig. 5B, in which the transmitted signals are affected significantly to reach flagrant errors.

In experiments, we fabricated two closely packed plasmonic waveguides, whose geometry parameters are the same as numerical simulations, based on the available printed circuit board technology, as shown in Figs. 6A and C. Here, the separation between two waveguides is chosen as $s$=0.8 mm (0.027 wavelengths at 10 GHz). For comparison, we also fabricated two closely packed microstrip lines with the same geometries (Figs. 6B and D). To conveniently measure the frequency spectra and near fields, we weld four standard SMA connectors to the four ports of samples. Using the S-parameters measurement technique introduced in *Materials and Methods*, we obtain the measured power transmission and coupling ratios from 0.03 to 20 GHz, and the results are presented in Fig. 7A, which have excellent agreements with the simulation results (Fig. 4A). We clearly notice that the plasmonic waveguides and microstrip lines have nearly the same transmission performance below the cutoff frequency. However, the crosstalk or mutual coupling between the two plasmonic waveguides can be effectively suppressed. Compared with the microstrip lines, the crosstalk in the plasmonic waveguides is 10 dB lower in average. Especially at the frequency of 11 GHz, the SPP crosstalk is decreased by 40 dB than the conventional way. That is to say, there is four-order magnitude difference between the crosstalks of plasmonic waveguide and microstrip, which can reduce the pressure of circuit designs and electromagnetic compatibility.

The measured time-domain transmitted SPP and microstrip signals are presented in Figs. 7B and C. We observe that the transmitted SPP signals are nearly distortion-less by passing through the plasmonic waveguide system due to the efficient interference suppression. However, the transmitted signals along microstrip lines have significant distortions. To show the crosstalk reduction in an ocular way, we also provide measurement results of near electric fields, as illustrated in Figs. 6E-H. The near-field experiments are conducted by using home-made equipment (*Materials and Methods*). We find that the spoof SPPs are propagating along their own waveguide with low loss with little mutual coupling to the other waveguide (Figs. 6E and G). As a contrast, the near-field distributions on the microstrip lines show longer wavelength and the more striking crosstalk (Figs. 6F and H). Hence, the interference suppressions and subwavelength effects of time-domain SPP signals are demonstrated experimentally.

## CONCLUSIONS

To break the challenges of signal integrity in integrated circuits and complex devices/systems, we have proposed time-domain spoof SPPs as the carrier of signals, which are supported by a double-strip plasmonic waveguide. When two plasmonic waveguides are tightly packed with deep-subwavelength separation, we have demonstrated that the SPP signals have much less mutual coupling than the conventional microstrip lines. The unique and beneficial features of strong field confinement, low crosstalk and natural filtration of the plasmonic waveguides have been verified by both frequency-spectrum measurements and near-field mappings. As one of the most potential candidates for SPP integrated circuits and systems, the proposed method is very attractive for a variety of applications, such as the high-speed circuits, parallel transmission systems, and conformal high-density integrated circuits. Meanwhile, the method can be extended to the higher frequencies (e.g., millimeter waves, terahertz waves, and far infrared).

## MATERIALS AND METHODS

### The S-parameter measurements

The key apparatus we used in the S-parameter measurements is an Agilent vector network analyzer (VNA N5230C). However, it cannot measure the 4-port system directly. To solve the problem, we select one of the ports as the input port connected to Port 1 of VNA and measure the transmission coefficient ($S_{21}$) and reflection coefficients ($S_{11}$) of other ports successively. Note that the ports except the port connected to VNA need to be loaded with the matching load over the measurements.

### The near-field measurements

We conduct the near-field measurements using a home-made near-electric-field mapper,

which is composed of a VNA (N5230C), a monopole antenna as the detector, and a planar platform which can move in the *x*- and *y*-directions automatically controlled by a stepper motor. The input port of the plasmonic waveguide is connected to Port 1 of VNA through SMA to feed the energy, and the other port is connected to the matched load to eliminate reflections. To probe the vertical (*z*) components of electric fields, the monopole antenna is fixed on top of the plasmonic waveguide around 1 mm and connected to Port 2 of VNA.

## SUPPLEMENTARY MATERIALS

Supplementary material for this article is provided.

**Fig. S1**. The geometry configuration of smooth conversions between the SPP waveguide and microstrip line.

**Fig. S2**. The eigen-mode results of SPP waveguides with (**A**) different phase shifts, and (**B**) different groove depths *d*.

**Fig. S3**. Simulation results of two closely-packed single-strip SPP waveguides, from which strong mutual coupling is observed. (**A**) The overall structure. (**B**) The simulated S-parameters. (**C**) The simulated time-domain signal spectra.

## REFERENCES AND NOTES

1.  H. Johnson and M. Graham, High-Speed Signal Propagation: Advanced Black Magic, Upper Saddle River, New Jersey: Prentice Hall, (2002).

2.  M. H. Nazari, A. E. Neyestanak, A 15-Gb/s 0.5-mW/Gbps two-tap DFE receiver with far-end crosstalk cancellation. *IEEE J. Solid-State Circuits* **47**, 2420–2432, (2012).

3.  S. H. Hall, L. H. Howard, advanced signal integrity for high-speed digital designs (Wiley-IEEE, 2009).

4.  W. L. Barnes, A. Dereux, T. W. Ebbesen, Surface plasmon subwavelength optics. *Nature* **424**, 824-830 (2003).

5.  H. Shin, S. H. Fan, All-Angle Negative Refraction for Surface Plasmon Waves Using a Metal-Dielectric-Metal Structure. *Phys. Rev. Lett.* **96**, 073907 (2006).

6.  L. L. Yin, et al. Subwavelength focusing and guiding of surface plasmons. *Nano Lett.* **5**,

1399-1402 (2005).

7. A. C. Jones, *et al.* Mid-IR Plasmonics: Near-Field Imaging of Coherent Plasmon Modes of Silver Nanowires. *Nano Lett.* **9**, 2553-2558 (2009).

8. N. Fang, H. Lee, C. Sun, X. Zhang, Sub–Diffraction-Limited Optical Imaging with a Silver Superlens. *Science* **308**, 534-537 (2005).

9. J. N. Anker, *et al.* Biosensing with plasmonic nanosensors. *Nature Mater.* **7**, 442-453 (2008).

10. A. Polman, H. Atwater, A. Photonic design principles for ultrahigh-efficiency photovoltaics. *Nature Mater.* **11**, 174-177 (2012).

11. N. Liu, *et al.* Individual Nanoantennas Loaded with Three-Dimensional Optical Nanocircuits. *Nano Lett.* **13**, 142-147 (2013).

12. A. P. Hibbins, B. R. Evans, J. R. Sambles, Experimental Verification of Designer Surface Plasmons. *Science* **308**, 670-672 (2005).

13. J. B. Pendry, L. Martin-Moreno, F. J. Garcia-Vidal, Mimicking surface plasmons with structured surfaces. *Science* **305**, 847-848 (2004).

14. F. J. Garcia-Vidal, L. Martin-Meoreno, J. B. Pendry, Surfaces with holes in them: new plasmonic metamaterials. *J.Opt. A Pure Appl Opt.* **7**, S97 (2005).

15. C. R. Williams, *et al.* Highly confined guiding of terahertz surface plasmon polaritons on structured metal surfaces *Nature Photonic.* **2**, 175-179 (2008).

16. S. A. Maier, S. R. Andrews, L. Martin-Moreno, F. J. Garcia-Vidal, Terahertz Surface Plasmon-Polariton Propagation and Focusing on Periodically Corrugated Metal Wires. *Phys. Rev. Lett.* **97** 176805 (2006).

17. A. A. Harry, P. Albert, Plasmonics for improved photovoltaic devices. *Nature Mater.* **9**, 205-213 (2010).

18. N. Prashant, C. L. Nathan, O. Sang-Hyun, J. N. David, Ultrasmooth Patterned Metals for Plasmonics and Metamaterials. *Science,* **325**, 594-597 (2009).

19. Y. J. Zhou, Q. Jiang, T. J Cui, Bidirectional bending splitter of designer surface plasmons. *Appl. Phys. Lett.* **99**,111904 (2011).

20. J. G. Rivas, Terahertz: The art of confinement. *Nature Photonic.* **2**, 137-138 (2008).

21. Q. Gan, Z. Fu, Y. J. Ding, F. J. Bartoli, Ultrawide-Bandwidth Slow-Light System Based on THz Plasmonic Graded Metallic Grating Structures. *Phys. Rev. Lett.* **100**, 256803 (2008).

22. X. Shen, T. J. Cui, D. Martin-Cano, F. J. Garcia-Vidal, Conformal surface plasmons propagating on ultrathin and flexible films. *P. Natl. Acad. Sci. USA.* **110**, 40-45 (2013).

23. H. C. Zhang, S. Liu, X. Shen, L. H. Chen, L. Li, T. J. Cui, Broadband amplification of spoof surface plasmon polaritons at microwave frequencies. *Laser & Photon. Rev.* **9**, 83-90 (2015).

24. J. Grgic´, J. R. Ott, F. Wang, O. Sigmund, A. Jauho, J. Mørk, N. Asger Mortensen, Fundamental Limitations to Gain Enhancement in Periodic Media and Waveguides. *Phys. Rev. Lett.* **108**, 183903 (2012).

25. Y. Yang, X. Shen, P. Zhao, H. C. Zhang, T. J. Cui, Trapping surface plasmon polaritons on ultrathin corrugated metallic strips in microwave frequencies. *Opt. Express* **23**, 7031–7037 (2015).

26. See Supplementary Material for detailed descriptions.

27. H. F. Ma, X. Shen, Q. Cheng, W. X. Jiang, T. J. Cui, Broadband and high-efficiency conversion from guided waves to spoof surface plasmon polaritons, *Laser & Photon. Rev.* **8**, 146–151 (2014)

28. S. Sun, Q. He, S. Xiao, Q. Xu, X. Li, L. Zhou, Gradient-index meta-surfaces as a bridge linking propagating waves and surface waves, *Nature Mater.* **11**, 426–431(2012).

29. X. Gao, L. Zhou, Z. Liao, H. F. Ma, T. J. Cui, An ultra-wideband surface plasmonic filter in microwave frequencies, *Appl. Phys. Lett.* **104**, 191603 (2014).

30. J. Y. Yin, J. Ren, H. C. Zhang, B. C. Pan, T. J. Cui, Broadband frequency-selective spoof surface plasmon polaritons on ultrathin metallic structure, *Sci. Rep.* **5**, 8165 (2013).

31. A. Manjavacas, F. J. García de Abajo, Robust plasmon waveguides in strongly interacting

nanowire arrays. *Nano Lett.* **9**, 1285–1289 (2009).

32. A. Manjavacas, F. J. García de Abajo, Coupling of gap plasmons in multi-wire waveguides. *Opt. Express* **17**, 19401–19413 (2009).

33. W. Cai, L. Wang, X. Zhang, J. Xu, F. J. Garcia de Abajo, Controllable excitation of gap plasmons by electron beams in metallic nanowire pairs. *Phys. Rev. B* **82**, 125454 (2010).

34. V. Myroshnychenko, *et al.* Interacting plasmon and phonon polaritons in aligned nano- and microwires. *Opt. Express* **20**, 10879–10887 (2012).

35. H. A. Haus, W. Huang, Coupled mode theory, *Proceedings of the IEEE* **19**, 1505-1518 (1991).

36. A. Ma, Y. Li, X. Zhang, Coupled mode theory for surface plasmon polariton waveguides, *plasmonics* **8**, 769–777 (2013).

**Acknowledgement:** H. C. Z. and T. J. C. contributed equally to this work. **Funding:** This work was supported by the National Science Foundation of China (61171024, 61171026, 61138001, 61302018 and 61401089), National High Tech (863) Projects (2012AA030402 and 2011AA010202), and the 111 Project (111-2-05). **Author contributions:** H. C. Z. proposed and analyzed the prototype of time-domain SPPs, implemented the simulations and measurements. T. J. C. proposed the concept and supervised the work. Q. Z. and Y. F. did the optimization and participated in measurements. X. F. contributed to the discussion. H. C. Z., X. F. and T. J. C. wrote the manuscript. **Competing interests:** The authors declare that they have no competing interests.

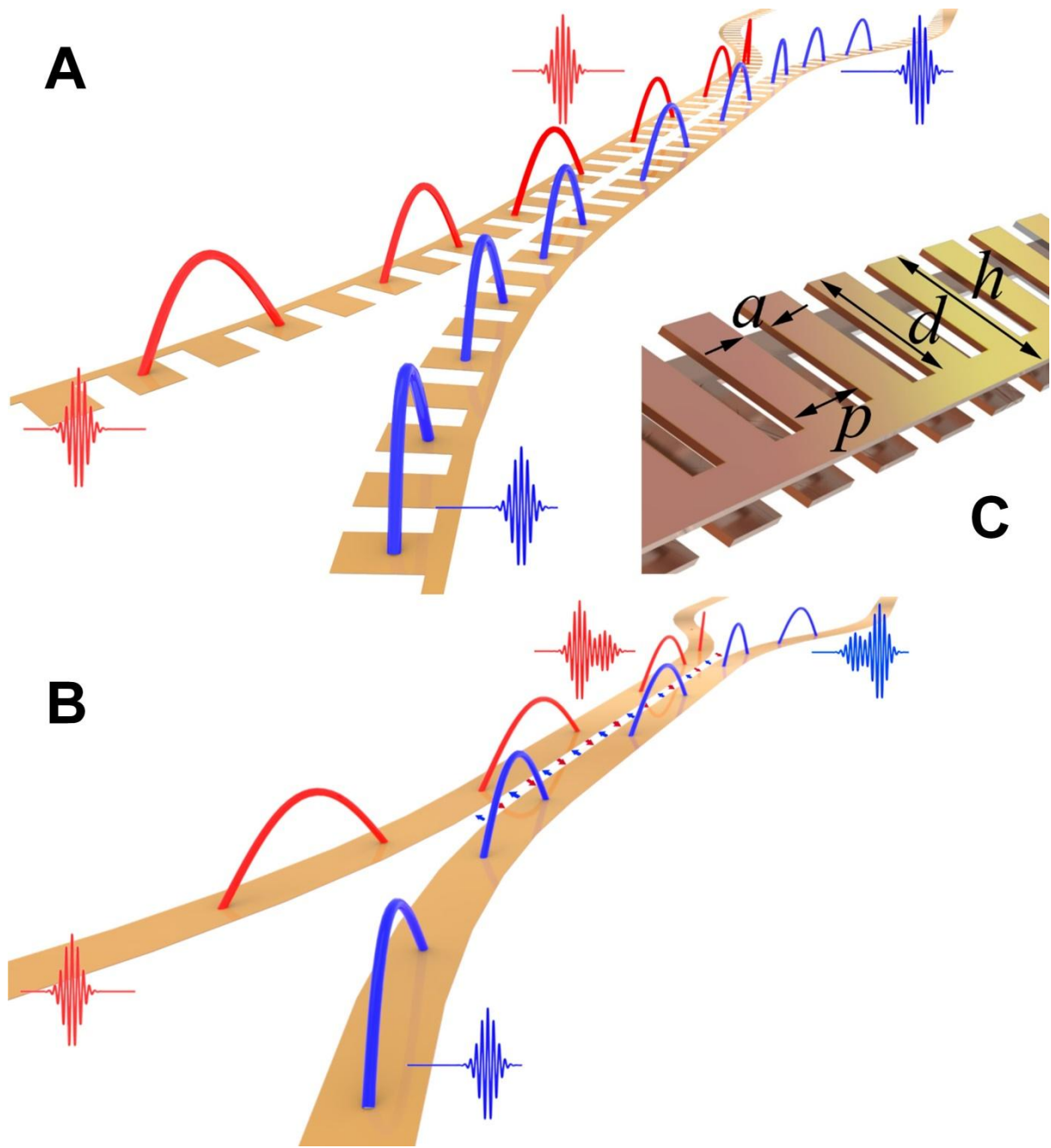

**Fig. 1. Two closely packed transmission lines with different types, which have different performance to suppress the interferences.** (**A**) Two closely packed spoof SPP waveguides, in which the interference is significantly suppressed. (**B**) Two closely packed microstrip lines, which have severe mutual interferences. (**C**) The detailed structure of the SPP waveguide.

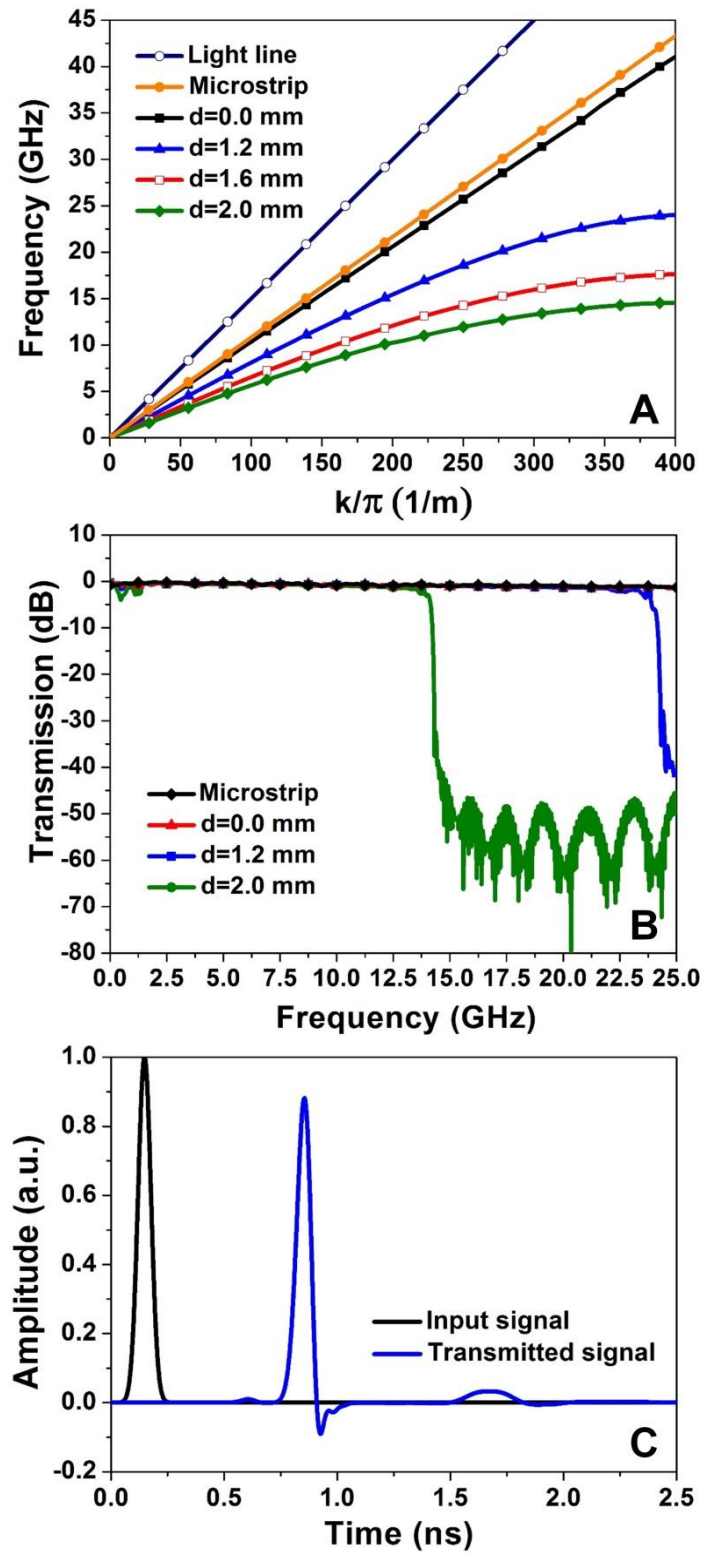

**Fig. 2. Simulated results of the spoof SPP waveguide, which can support time-domain SPP signals.** (**A**) Dispersion diagrams for different groove depths. (**B**) Transmission spectra of transmission lines with different types. (**C**) The time-domain SPP signal under the input of a Gaussian pulse with broad band from 0 to 12 GHz.

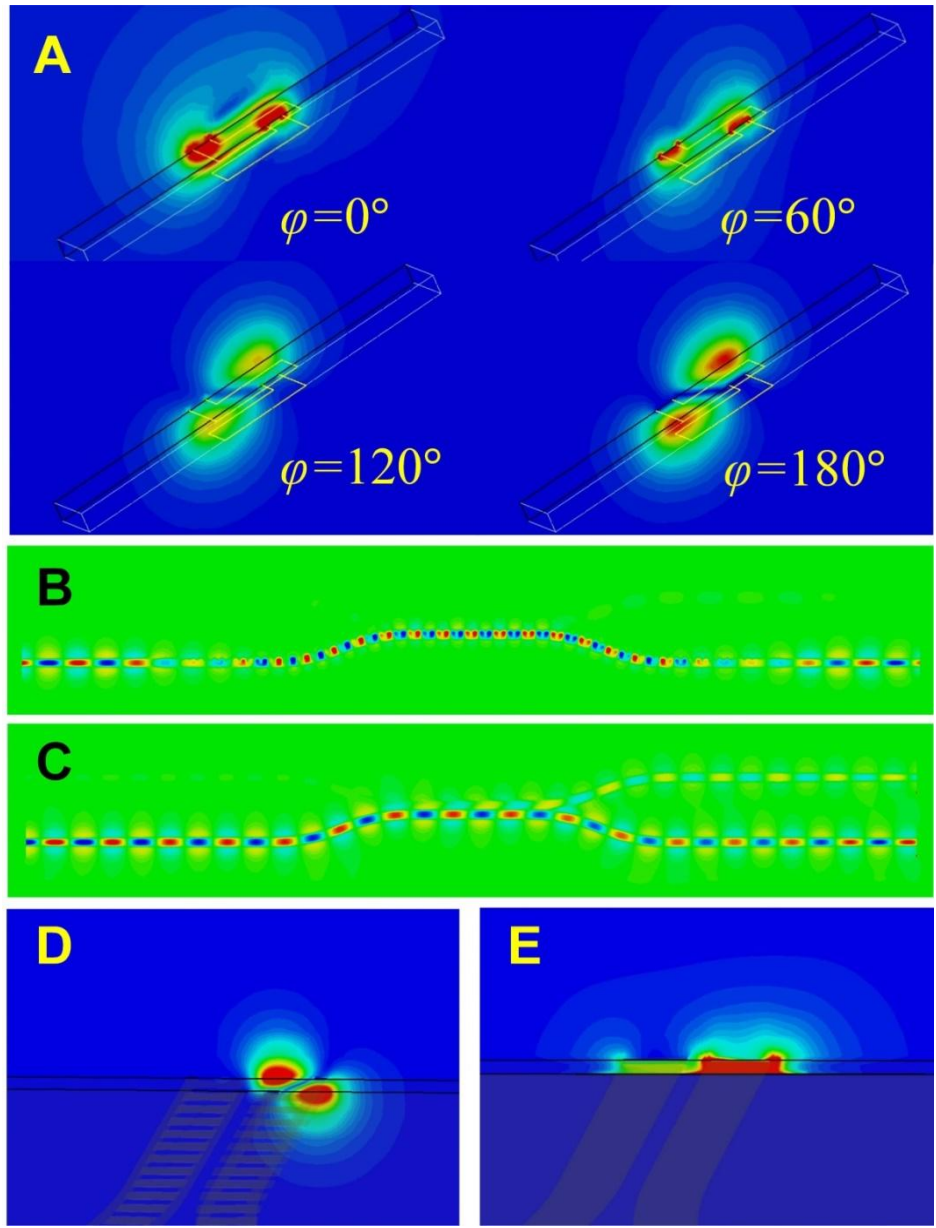

**Fig. 3. Simulated near electric-field distributions of the spoof SPP waveguide in different cases.** (**A**) The magnitude distributions of Eigen modes with different phase shifts (Color bar: from 0 to 2410 V/m). (**B**) The normal electric-field distribution of two closely-packed SPP waveguides on the *x-y* plane that is 1.0 mm above the plasmonic structure at 10 GHz (Color bar: from -1500 to -1500 V/m). (**C**) The normal electric-field distribution of two closely packed microstrip lines on the same observation plane at 10 GHz (Color bar: from -1500 to -1500 V/m). (**D**) The magnitude distributions of two closely-packed SPP waveguides on the *y-z* plane at 10 GHz (Color bar: from 0 to 2300 V/m). (**E**) The magnitude distributions of two closely-packed microstrip lines on the *y-z* plane at 10 GHz (Color bar: from 0 to 2300 V/m).

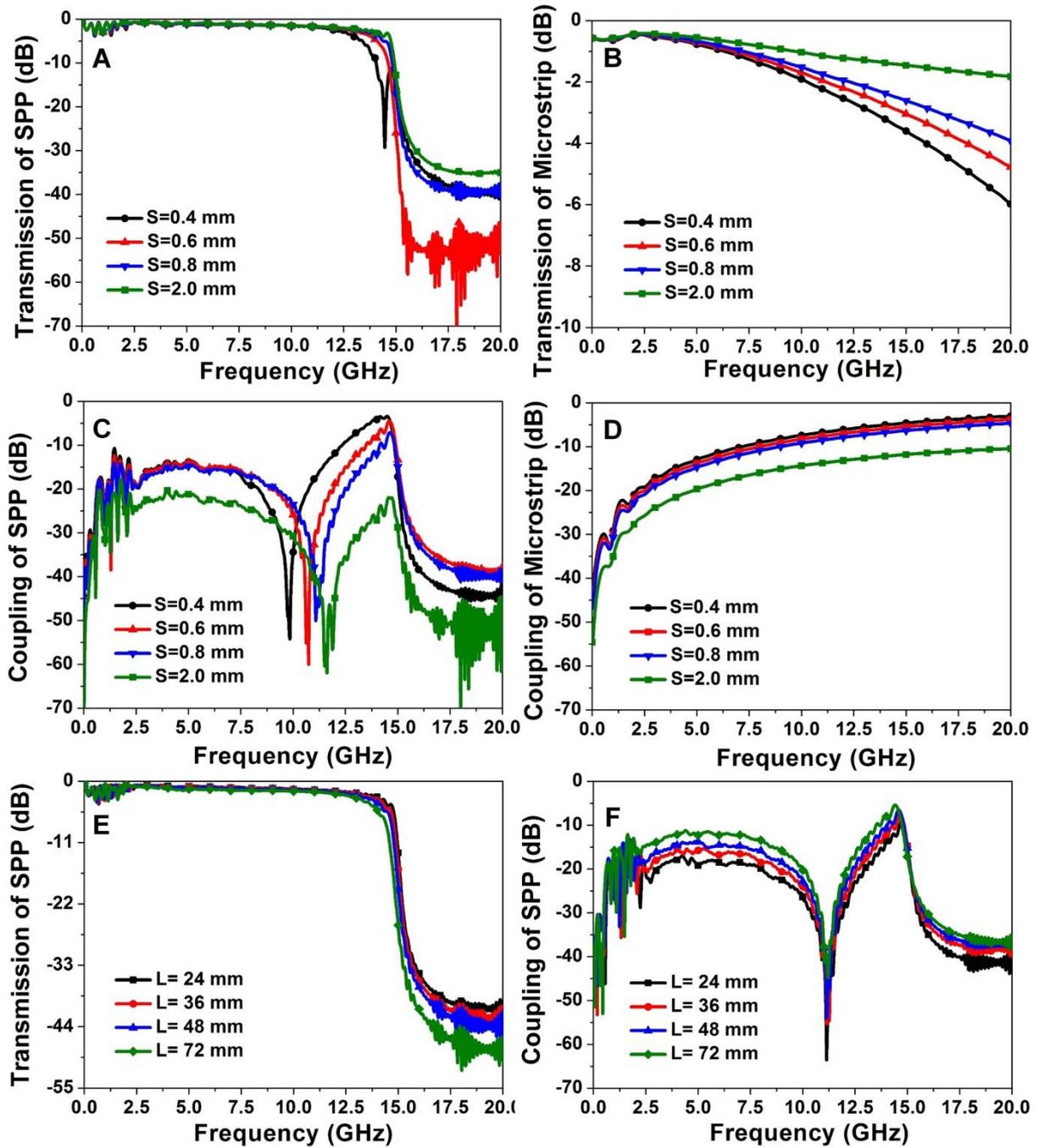

**Fig. 4. Simulated transmission and coupling coefficients in different cases.** (**A** and **B**) The transmission coeffcientss of two closely-packed SPP waveguides (**A**) and microstrip lines (**B**) with different separations. (**C** and **D**) The coupling of two closely-packed SPP waveguides (**C**) and microstrip lines (**D**) with different separations. (**E** and **F**) Transmission (**E**) and coupling (**F**) coefficients of two closely-packed SPP waveguides with different lengths.

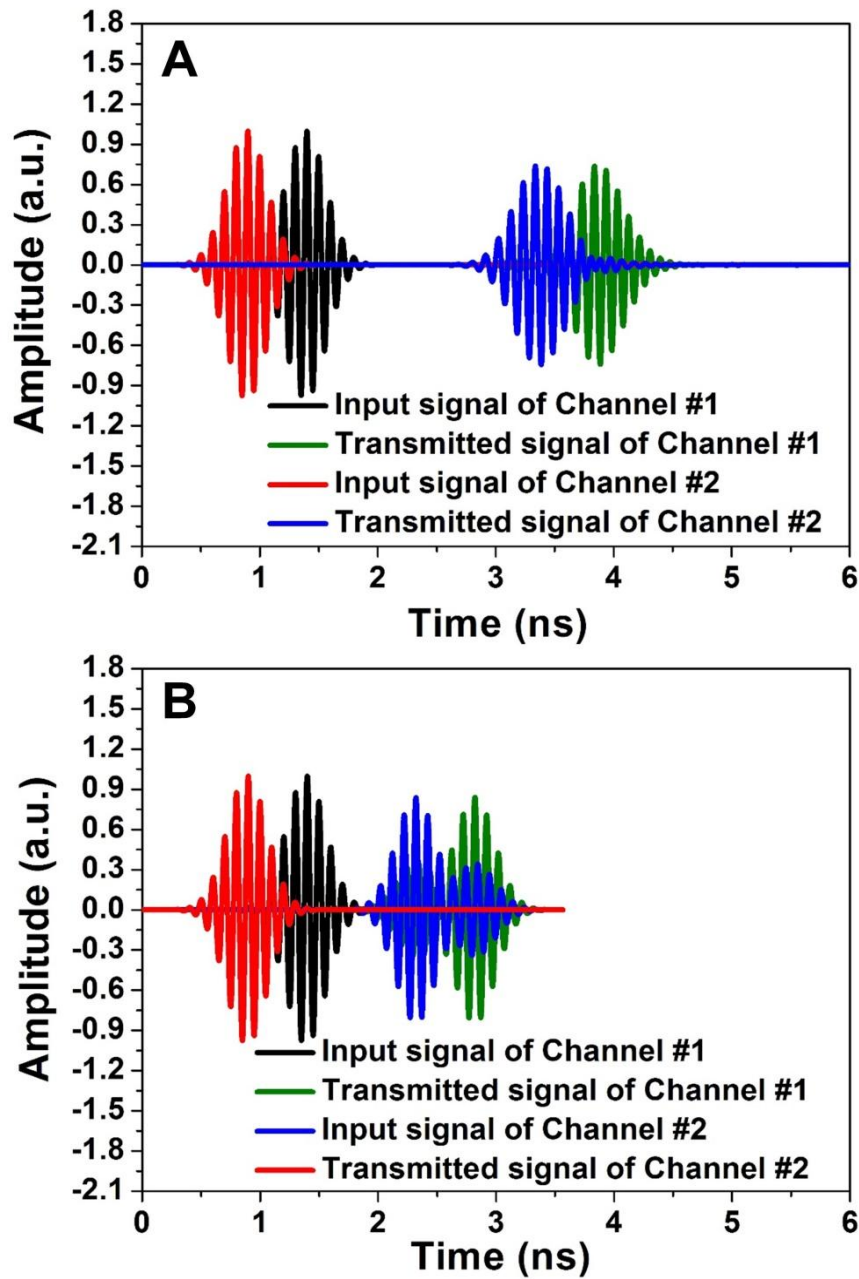

**Fig. 5. Simulated time-domain signal spectra of two closely packed transmission lines under the inputs of broadband Gaussian pulses (8-12 GHz) with 0.5 ns time difference.** (**A**) Two closely-packed SPP waveguides. (**B**) Two closely-packed microstrip lines.

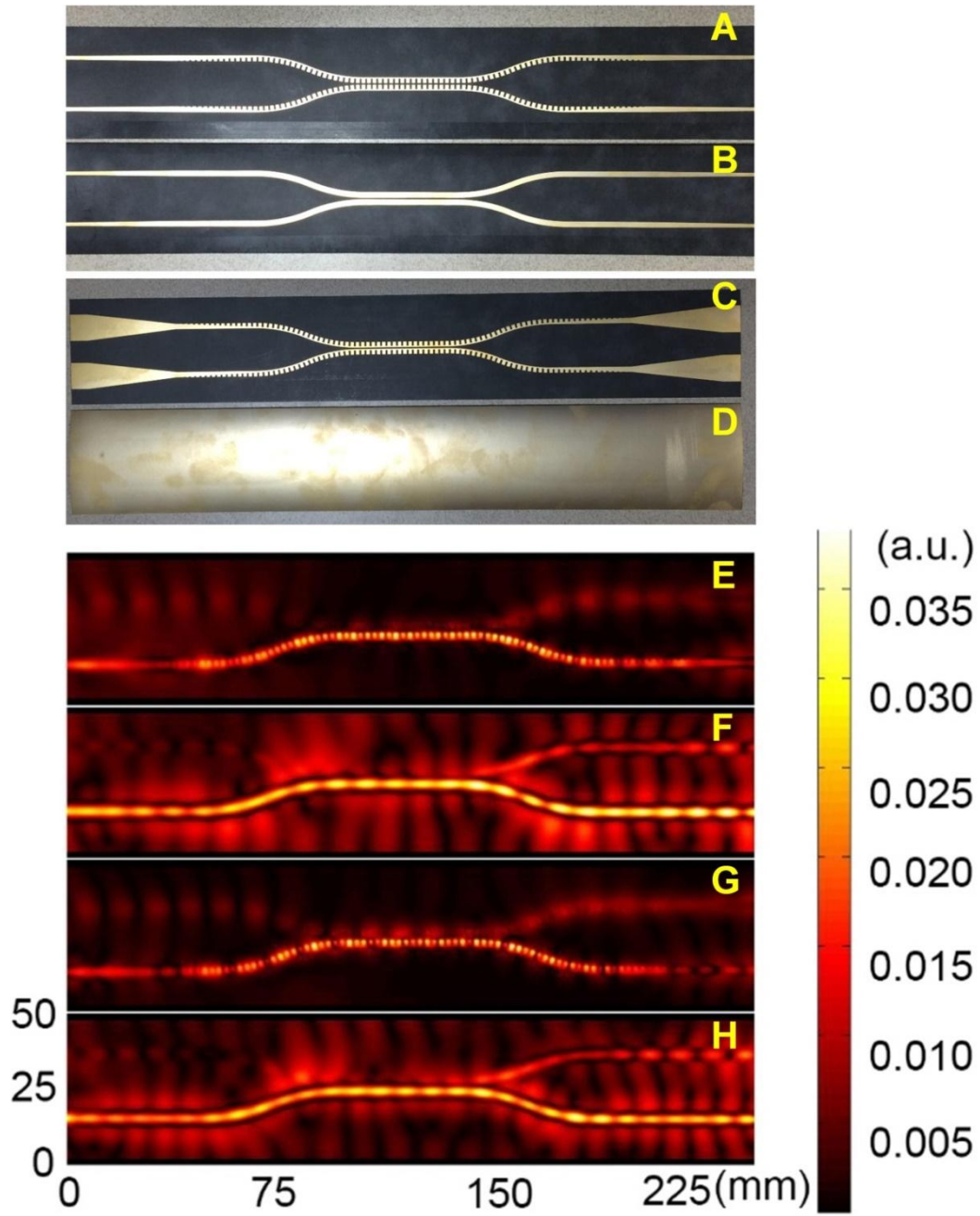

**Fig. 6. The fabricated samples and measured near-field results of the two closely-packed transmission lines.** (**A** and **B**) The top views of two closely-packed SPP waveguides (**A**) and microstrip lines (**B**). (**C** and **D**) The bottom views of two closely-packed SPP waveguides (**C**) and microstrip lines (**D**). (**E** and **F**) The measured near-field mapping results of the two closely-packed SPP waveguides (**E**) and microstrip lines (**F**) at 11 GHz. (**G** and **H**) The measured near-field mapping results of the two closely-packed SPP waveguides (**G**) and microstrip lines (**H**) at 12.5 GHz.

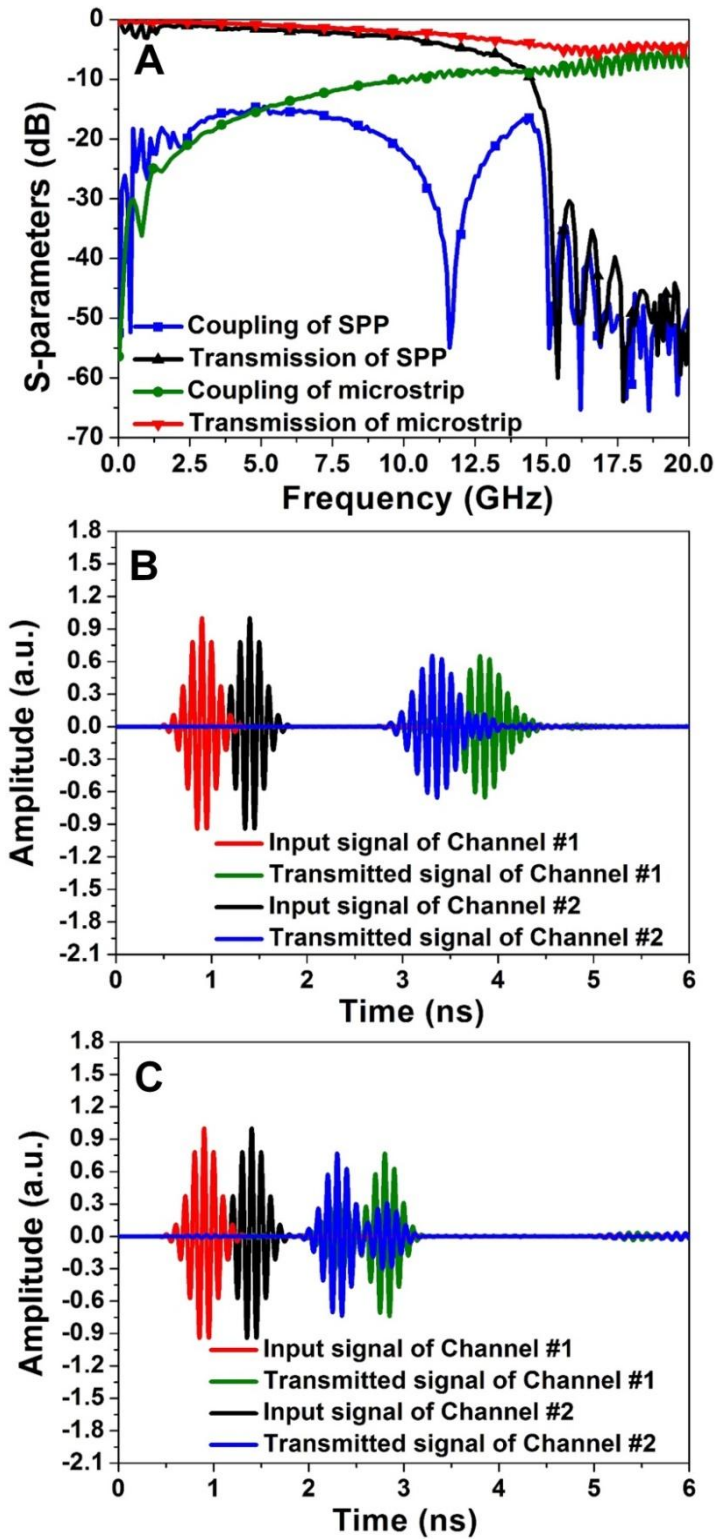

**Fig. 7. The measured result of two closely-packed transmission lines.** (**A**) The measured transmission and coupling coefficients of two closely-packed SPP waveguides and microstrip lines. (**B** and **C**) The time-domain signal spectra of two closely-packed SPP waveguides (**B**) and microstrip lines (**C**) under the inputs of broadband Gaussian pulses (8-12 GHz) with 0.5 ns time difference.

# Supplementary Materials for

# Breaking the challenge of signal integrity using time-domain spoof surface plasmon polaritons

Hao Chi Zhang[1,2,*], Tie Jun Cui[1,3,*†], Qian Zhang[1,2], Yifeng Fan[1,2], and Xiaojian Fu[1,2]

[1] *State Key Laboratory of Millimeter Waves, Southeast University, Nanjing 210096, China*
[2] *Synergetic Innovation Center of Wireless Communication Technology, Southeast University, Nanjing 210096, China*
[3] *Cooperative Innovation Centre of Terahertz Science, No.4, Section 2, North Jianshe Road, Chengdu 610054, China*

*These authors contributed equally to this work.
†To whom correspondence should be addressed. E-mail: tjcui@seu.edu.cn

**This PDF file includes:**

The detailed descriptions of 1) smooth conversions between the plasmonic waveguide and microstrip line; 2) eigen modes and excited modes of the plasmonic waveguide; and 3) numerical results of two closely-packed single-strip SPP waveguides.

**Fig. S1**. The geometry configuration of smooth conversions between the SPP waveguide and microstrip line.

**Fig. S2**. The eigen-mode results of SPP waveguides with (**A**) different phase shifts, and (**B**) different groove depths *d*.

**Fig. S3**. Simulation results of two closely-packed single-strip SPP waveguides, from which strong mutual coupling is observed. (**A**) The overall structure. (**B**) The simulated S-parameters. (**C**) The simulated time-domain signal spectra.

**Smooth conversions between the plasmonic waveguide and microstrip line**

Aimed to achieve the transition from the conventional microstrip line to plasmonic waveguide, two structures have been proposed *(1-2)*, which can efficiently motivate the SPP modes by the microstrip line. In consideration of the feature of double-strip structure (the right subfigure in Fig. S1), we have improved the design in Ref. *(2)* and optimize all geometrical parameters to realize smooth conversions between the plasmonic waveguide and microstrip, in which the geometrical parameters are designed as: $L_1$=8 mm, $L_2$=40 mm, $L_3$=40 mm, $w_1$=1.588 mm and $w_2$=12 mm, as shown in Fig. S1.

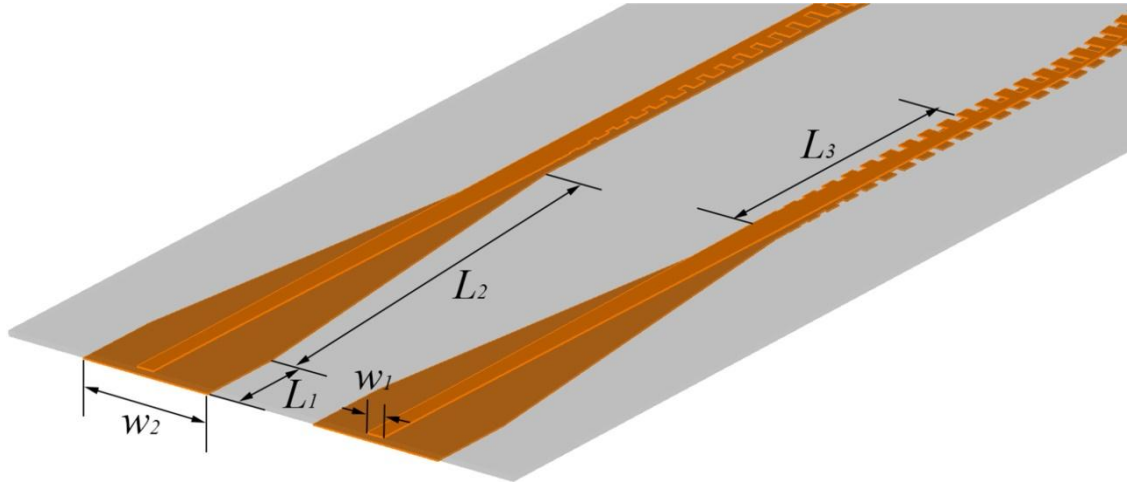

**Fig. S1**. The geometry configuration of smooth conversions between the SPP waveguide and microstrip line.

### Eigen modes and the excited modes of the plasmonic waveguide

For clearly showing the electric distribution of spoof SPPs, we provide the front-view figure of eigen modes (Fig. S2A). We notice that the field distribution is transposed from the gap SPP mode to surface-wave mode. This special phenomenon is caused by the enhanced and confinement abilities of the double-strip plasmonic waveguide. According to the dispersion curve of spoof SPPs, the confinement ability can also be tuned by changing the geometry parameters (e.g. groove depths *d*). Hence we can realized the gap SPP mode and SPP mode at the required frequencies, as shown in Fig. S2B, which can be utilized to excite the SPP mode. Compared to the SPP mode, the smooth conversions between the gap SPP mode and spatial mode in microstrip are more easily to achieve due to the similar field distributions in the plane perpendicular to the propagating direction.

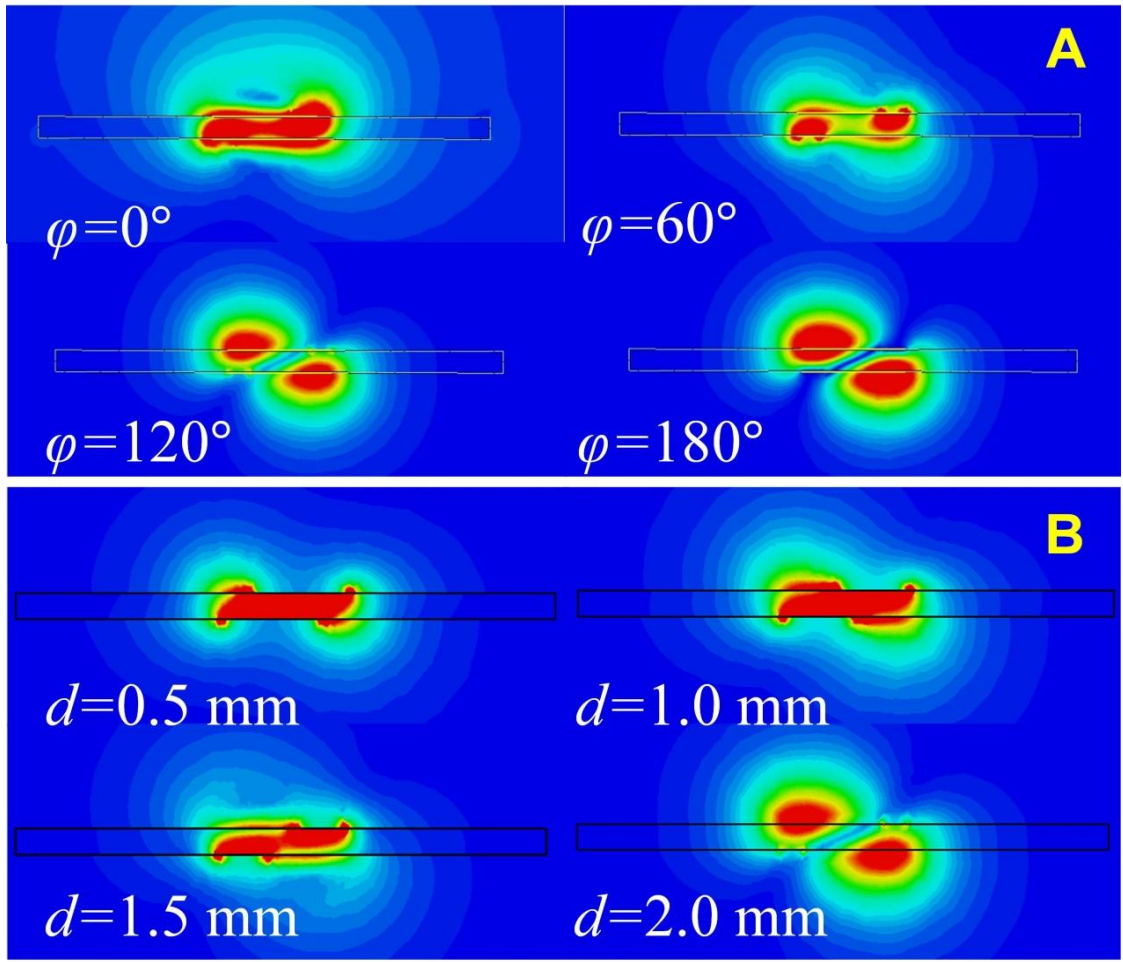

**Fig. S2**. The eigen-mode results of SPP waveguides with (**A**) different phase shifts, and (**B**) different groove depths $d$.

**Numerical results of two closely-packed single-strip SPP waveguides**

As one of most potential candidates of conformal circuits, conformal surface plasmons (CSPs) *(3)* and related functional devices *(4-9)* on single-strip waveguides have been intensively studied in recent years. Hence we select single-strip SPP waveguides as the other comparison, as shown in Fig. S3A. The numerical results of S-parameters and time-domain spectra for two closely-packed single-strip SPP waveguides are illustrated in Figs. S3B and S3C, respectively. We clearly observe that the adjacent waveguides behave like a spoof SPP coupler *(9)* rather than two non-interfering transmission lines. That is to say, the coupling coefficient $\kappa$ of the single-strip structure is much larger than that of double-strip structure, which cannot be used to suppress the interactions between the two SPP waveguides.

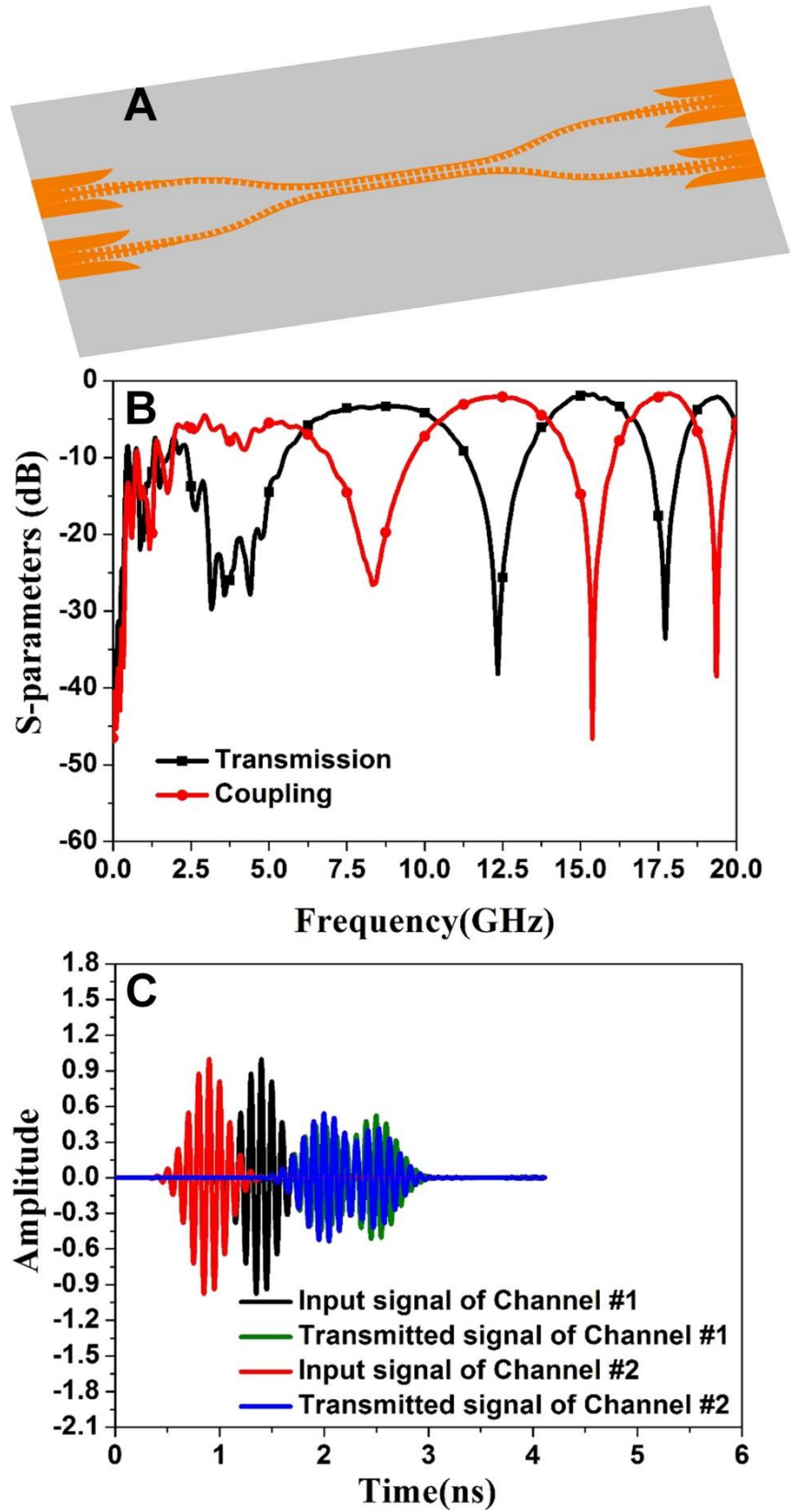

**Fig. S3**. Simulation results of two closely-packed single-strip SPP waveguides, from which strong mutual coupling is observed. (**A**) The overall structure. (**B**) The simulated S-parameters. (**C**) The simulated time-domain signal spectra.

# REFERENCES AND NOTES

1. Z. Liao, J. Zhao, B. C. Pan, X. P. Shen, T. J. Cui, Broadband transition between microstrip line and conformal surface plasmon waveguide, *J. Phys. D: Appl. Phys.* **47**, 315103 (2014).

2. H. C. Zhang, S. Liu, X. Shen, L. H. Chen, L. Li, T. J. Cui, Broadband amplification of spoof surface plasmon polaritons at microwave frequencies. *Laser & Photon. Rev.* **9**, 83-90 (2015).

3. X. Shen, T. J. Cui, D. Martin-Cano, F. J. Garcia-Vidal, Conformal surface plasmons propagating on ultrathin and flexible films. *P. Natl. Acad. Sci. USA.* **110**, 40-45 (2013).

4. X. Gao, L. Zhou, Z. Liao, H. F. Ma, T. J. Cui, An ultra-wideband surface plasmonic filter in microwave frequencies, *Appl. Phys. Lett.* **104**, 191603 (2014).

5. J. Y. Yin, J. Ren, H. C. Zhang, B. C. Pan, T. J. Cui, Broadband frequency-selective spoof surface plasmon polaritons on ultrathin metallic structure, *Sci. Rep.* **5**, 8165 (2013).

6. X. Shen, and T. J. Cui, Planar plasmonic metamaterial on a thin film with nearly zero thickness, *Appl. Phys. Lett.* **102**, 211909 (2013).

7. J. J. Xu, H. C. Zhang, Q. Zhang, T. J. Cui, Efficient conversion of surface-plasmon-like modes to spatial radiated modes, *Appl. Phys. Lett.* **106**, 021102 (2015).

8. B. C. Pan, Z. Liao, J. Zhao, T. J. Cui, Controlling rejections of spoof surface plasmon polaritons using metamaterial particles, *Opt. Express* **22**, 13940 (2014).

9. X. Liu, Y. Feng, K. Chen, B. Zhu, J. Zhao, T. Jiang, Planar surface plasmonic waveguide devices based on symmetric corrugated thin film structures, *Opt. Express* **22**, 20107 (2014).